\documentclass[onecolumn,showpacs,superscriptaddress,amsmath,amssymb,pre]{revtex4}

\usepackage{graphicx}
\usepackage{dcolumn}
\usepackage{bm}

\newcommand*{\beq}{\begin{equation}}
\newcommand*{\eeq}{\end{equation}}
\newcommand{\Teleia}{\quad .}
\newcommand{\Comma}{\quad ,}
\newcommand*{\vp}{\bm{v}}
\newcommand*{\uf}{\bm{u}}
\newcommand*{\vprime}{\bm{v}^{\prime}}

\newcommand*{\vmeanb}{\overline{\bm{v}}}

\newcommand*{\xvec}{\bm{x}}
\newcommand*{\tprime}{t^{\prime}}
\newcommand*{\boldnabla}{\mbox{\boldmath$\nabla$}}
\newcommand*{\Stk}{\textrm{St}}

\begin{document}

\title{Brownian motion of finite-inertia particles in a simple shear flow}

\author{Yannis Drossinos}
\email{ioannis.drossinos@jrc.it} \affiliation{European Commission,
Joint Research Centre, I-21020 Ispra (Va), Italy}
\author{Michael W. Reeks}
\email{Mike.Reeks@newcastle.ac.uk} \altaffiliation{Permanent
address: School of Mechanical and Systems Engineering, University
of Newcastle, Newcastle upon Tyne, NE1 7RU United Kingdom.}
\affiliation{European Commission, Joint Research Centre, I-21020
Ispra (Va), Italy}

\date{\today}

\begin{abstract}
Simultaneous diffusive and inertial motion of Brownian
particles in laminar Couette flow is investigated via Lagrangian
and Eulerian descriptions to determine the effect of
particle inertia on diffusive transport in the
long-time limit. The classical fluctuation
dissipation theorem is used to calculate the amplitude of
random-force correlations, thereby neglecting corrections of the
order of the molecular relaxation time to the inverse shear rate.
In the diffusive limit (time much greater than the particle
relaxation time) the fluctuating particle-velocity
autocorrelations functions are found to be stationary in time, the
correlation in the streamwise direction being an exponential multiplied by an
algebraic function and the cross correlation non-symmetric in the
time difference. The analytic, non-perturbative, evaluation of the
particle-phase total pressure, which is calculated to be second
order in the Stokes number (a dimensionless measure of particle
inertia), shows that the particle phase behaves as a non-Newtonian
fluid. The generalized Smoluchowski convective-diffusion equation,
determined analytically from a combination of the
particle-phase pressure tensor and the inertial acceleration term,
contains a shear-dependent cross derivative term and an
additional term along the streamwise direction,
quadratic in the particle Stokes number. The long-time diffusion coefficients
associated with the particle flux relative to the carrier flow
are found to depend on particle inertia such that the streamwise diffusion
coefficient becomes negative with increasing
Stokes number, whereas one of the cross coefficients is always
negative. The total diffusion coefficients
measuring the rate of change of particle mean square
displacement are always positive as expected
from general stability arguments.
\end{abstract}


\maketitle

\section{Introduction}
\label{sec:intro}

Combined inertial and diffusive motion of Brownian particles in a
flowing fluid is important in a number of aerosol processes, including
filtration, aerosol sampling, deposition in bends, and particulate
deposition in the human respiratory tract. The limit of negligible
inertial effects, where the particles follow closely the motion of the fluid,
has been extensively studied in sheared colloidal
suspensions~\cite{ronis84,hess84}. These two-phase systems, which consist of a dispersed
particulate phase and a continuous fluid phase (gas for aerosols),
have long been of interest for their important industrial and
engineering applications. Pure diffusive particle motion is usually
described by a convective gradient-diffusion equation
from which inertial effects are absent. The limit of inertial
transport, where diffusion is neglected and particle trajectories deviate
significantly from the fluid streamlines, is most conveniently described in
terms of the particle equations of motion in a Lagrangian formulation.
In the transition regime between the diffusion limit and the
inertia-dominated limit the two particle-transport mechanisms have to be
considered simultaneously.

The effect of particle inertia on the diffusive motion of non-interacting
Brownian particles in nonuniform fluids has been examined via numerous
approaches. Continuum descriptions in terms of mass and momentum
conservation equations (macroscopic ``hydrodynamic'' equations)
require constitutive relations for the particle-phase total pressure tensor.
A frequently made
approximation is to consider the Brownian particles as an ``ideal gas'', thereby
using a phenomenological expression for the particle-phase
pressure tensor~\cite{ramshaw79,delaMora82}.

Mesoscopic descriptions of Brownian motion involve stochastic
particle equations of motion and the associated Fokker-Planck
equation, as, for example, in Subramanian and Brady~\cite{brady} where a multiple
scale analysis of the Fokker-Planck equation in a simple shear
flow is presented. In mesoscopic descriptions in terms of Langevin
equations the fluctuation dissipation theorem (FDT) is an
essential ingredient of the calculation. Santamar\'{i}a-Holek et
al.~\cite{rubi01} used an alternative approach by considering the
motion of a Brownian particle in an external flow field as an example
of a driven, far-from-equilibrium system. They used mesoscopic
nonequilibrium thermodynamics, an approach that does not require
the specification of the stochastic properties of the random
force, to obtain the Fokker-Planck equation for the nonequilibrium
Brownian particle distribution function in a simple shear flow.
They found that the diffusion tensor in the Fokker-Planck equation
depends on the shear rate, concluding that fluctuations about the
nonequilibrium steady state lead to a violation of the classical
(equilibrium) FDT.

Kinetic theory has also been used to obtain the Fokker-Planck equation for
the motion of Brownian particles in rarefied nonuniform gases.
Fern\'{a}ndez de la Mora and Mercer~\cite{delaMorapra82b} expanded the
Boltzmann collision operator in the ratio of the light-gas molecular mass to
the Brownian-particle mass; they approximated the light-gas distribution
function by the first two terms in the Chapman-Enskog expansion to derive a
Fokker-Planck equation whose diffusion tensor was found independent of
light-gas velocity gradients~\cite{delaMorapra82b}. However, Rodr\'{i}gez et
al.~\cite{dufty83}, using a similar low mass-ratio expansion of the
collision operator for Maxwellian fluid molecules, obtained a
Fokker-Planck equation in a uniform shear flow with a diffusion tensor that
depended on the irreversible fluid stress tensor.

Herein, we consider a dilute suspension of non-interacting Brownian
particles in a two-dimensional, simple shear (laminar, plane Couette flow). Inertial
Brownian particles are considered, namely particles whose Stokes number
(the ratio of the particle relaxation time to the inverse
shear rate, or more properly strain rate, a dimensionless number that describes
particle response to spatial changes in the carrier flow velocity) is
finite (at least of order unity). The limit of small
Stokes number corresponds to pure
diffusive motion, whereas for large Stokes numbers inertial transport
dominates.
A mesoscopic approach is adopted, whereby both Lagrangian and
Eulerian descriptions are presented to investigate the effect of particle
inertia on the convective-diffusion equation (equivalently, the Smoluchowski
equation) in the long-time, diffusive limit.
The classical FDT is used since we argue that the
modification of the classical FDT derived in Refs.~\cite{rubi01,dufty83} is
of the order of the ratio of the molecular (fluid) relaxation time (a
molecular time scale) to the inverse shear rate (a macroscopic time scale),
and thus negligible for typical shear rates.

A linear flow field was chosen as the underlying
carrier-gas velocity field
because the stochastic particle equations of motion and
the associated Fokker-Planck equation can be solved exactly
for linear flow fields. Thus, the closure problem
associated with the continuum equations for the dispersed phase
is avoided, and analytic, non-perturbative
expressions may be derived in the long-time limit
(for example, for the mean particle velocity).
The specific case of a simple shear
is investigated because Brownian motion in a simple
shear has been studied extensively, see, for example,
Refs.~\cite{brady,rubi01,dufty83,sanmiguel}. These
analyses have been limited in either
perturbative calculations~\cite{brady,rubi01},
or in calculations of the long-time behavior of equal-time correlation
functions without considerations of the associated
convective-diffusion equation~\cite{sanmiguel}.
The long-time, analytic expressions obtained herein extend
these previous analyses (under
well-specified approximations):
for example, the non-perturbative, generalized Smoluchowski
equation for diffusing, inertial
Brownian particles in a simple shear is derived,
see Ref.~\cite{brady}, and analytic expressions for
the particle-phase total pressure tensor and the diffusion tensor
are obtained, see Ref.~\cite{rubi01}.
Moreover, the methodology presented for a simple shear
may be used to investigate Brownian motion of finite-inertia
particle in any other linear two-dimensional flow,
be it a symmetric or antisymmetric (rotational) shear.

Our approach has much in common with the so-called PDF approach used to
describe particle dispersion in inhomogeneous turbulent flows~\cite{mike92}.
Moreover, a Langevin equation formally equivalent to the one used in this
work has also been used to model turbulence~\cite{pope} by considering
an analogy between the action of the dissipating
scales of turbulence and that of the molecular white noise driving force
in Brownian motion.
Hence, the results obtained herein apply \textit{mutatis
mutandis} for turbulent dispersion in a simple shear with the proviso that
FDT is not a property of the turbulent motion. Similarly, a Langevin
equation equivalent to the one presented in this work for the fluctuating
Brownian particle velocities has also been used to describe the effect of
shear on fluid velocity fluctuations~\cite{eckhardt03}. In the
identification of similarities and differences between Brownian-particle
motion and fluid-point motion the Stokes number used in this work
corresponds to the dimensionless
ratio of the fluid time scale to the inverse shear rate.

The Lagrangian description is presented in Sec.~\ref{sec:lagrangian} where
the Brownian-particle velocity autocorrelation functions are derived.
Section~\ref{sec:eulerian} contains the Eulerian description and the
analytic solution of the Fokker-Planck equation. The analytic expression for
the total particle-phase pressure tensor as a function of the Stokes number
is derived. The generalized Smoluchowski equation is calculated in
Sec.~\ref{sec:smoluchowski}, as well as the Green-Kubo expressions for the diffusion
coefficients. The conclusions are summarized in Sec.~\ref{sec:conclusions},
whereas technical details are presented in the Appendix.

\section{Lagrangian description}

\label{sec:lagrangian}

Consider, thus, a Brownian particle of mass $m$ in a two-dimensional
unbounded laminar, plane Couette shear flow, namely a simple shear flow with
the fluid velocity $\uf$ along the $y$-direction, $\uf =\tensor{\bm{\alpha}}%
\xvec$; the shear rate~\footnote{Strictly speaking
$\alpha $ is the strain rate, but we will follow the
standard practice of referring to it as the shear rate.}
tensor $\tensor{\bm{\alpha}}$ has only one non-zero
element, $\alpha _{xy}=\alpha$.
The general equation of motion for a small rigid sphere in a
nonuniform flow~\cite{maxey83} simplifies considerably for most
incompressible gas-particle systems because particle
density is much greater than fluid density. Accordingly,
the pressure gradient force, virtual mass, Basset history integral, and
Faxen's modification to Stokes' drag may be neglected. Moreover, the effect
of gravity will be neglected since gravitational settling becomes
significant only for very large particles ($r_p \gg 50 \mu m$). The Saffman
lift force will also be neglected, it being negligible with
respect to Stokes drag for small-diameter, low-inertia particles.
Therefore, the particle equations of motion in a Lagrangian description
become
\beq
\frac{d\vp}{dt} = \beta (\tensor{\bm{\alpha}} \xvec - \vp ) + \bm{f}(t)
\quad ,
\label{eq:eom}
\eeq
where the time derivative is a total derivative following
the moving Brownian particle
with $\vp(t)$ the particle velocity, $\xvec(t)$ the particle position, and $%
\bm{f}(t)$ the random force per unit particle mass. As argued, the friction force is
assumed to be the Stokes drag on the particle.
Hence, the friction coefficient $\beta $ is the inverse particle relaxation time,
$\beta =1/\tau _{p}=9\mu _{f}/(2\rho _{p}r_{p}^{2})$, the particle relaxation
time being $\tau _{p}$, the particle material density being $\rho _{p}$, the
particle radius $r_{p}$, and the fluid dynamic viscosity $\mu _{f}$ . As in
the Langevin description of Brownian motion in a quiescent fluid the random
force will be taken to be 
white in time
\beq
\langle f_{i}(t) f_{j}(\tprime) \rangle = q \delta _{ij}\delta (t-\tprime)
\quad \quad (i,j=x,y)
\quad ,
\label{eq:ff}
\eeq
with zero mean $\langle f_{i}(t)\rangle =0$ and of an unspecified, at the
moment, strength $q$. Angular brackets $\langle \cdot \rangle $ denote an
ensemble average over all particle trajectories.

The Langevin equations are linear: hence, the particle velocity and position
may be solved formally as functionals of the random force. The formal
solution for Brownian particles injected at the origin with zero initial
velocity (the choice of the initial conditions is not important since we are
interested in the long-time behavior) are
\begin{subequations}
\begin{align}
v_x(t) & = e^{-\beta t} \, \int_0^t \, dt_1 \, e^{\beta t_1}
\left [ \beta \alpha y(t_1) + f_x(t_1) \right ]
\quad & ; \quad
x(t) = \int_0^t \, dt_1 \, v_x(t_1)
\Comma \\
v_y(t) & = e^{-\beta t}
\int_0^t \, dt_1 \, e^{\beta t_1} f_y(t_1)
\quad & ; \quad
y(t) = \int_0^t \, dt_1 \, v_y(t_1)
\Teleia
\label{eq:solveB}
\end{align}
\label{eq:solve}
\end{subequations}
The formal solution allows the analytic evaluation of ensemble averages of
products of particle position and velocity, two-point correlation functions,
in terms of the random-force strength $q$. Their evaluation requires the
calculation of time integrals whose integrands include the causal
correlation
\beq
\langle y(t) f_y(\tprime) \rangle = \theta (t-\tprime) \frac{q}{\beta}
\left [ 1 - e^{-\beta(t-\tprime)} \right ] \quad .
\eeq
The Heaviside $\theta$ function in the previous equation arises
naturally via the explicit evaluation of the time-dependent
correlation function using Eqs.~\ref{eq:solveB}; it ensures causality.

In the diffusive limit ($t \gg \beta ^{-1}$) exponential terms in the
equal-time, two-point correlation functions may be neglected. If only
polynomial terms in time are kept the correlations evaluate to
\begin{subequations}
\beq
\langle \xvec(t) \xvec(t) \rangle = \frac{q}{2 \beta^2}
\begin{pmatrix}
2 t - \frac{3}{\beta} +
\alpha^2 \left [ \frac{2}{3} t^3 - \frac{4 t^2}{\beta}
+ \frac{8 t}{\beta^2} - \frac{3}{2 \beta^3} \right ] \quad \quad \quad
&
\alpha \left [ t^2 - \frac{4 t}{\beta} + \frac{11}{2 \beta^2} \right ] \\
\alpha \left [ t^2 - \frac{4 t}{\beta} + \frac{11}{2 \beta^2} \right ] &
2 t - \frac{3}{\beta}
\end{pmatrix}
\Comma
\label{eq:xx}
\eeq
\beq
\langle \bm{v}(t) \xvec(t) \rangle = \frac{q}{2 \beta^2}
\begin{pmatrix}
\alpha^2 \left [ t^2 - \frac{4 t}{\beta} + \frac{4}{\beta^2} \right ] + 1
\quad \quad \quad &
\alpha \left [ 2t - \frac{9}{2 \beta} \right ] \\
\frac{\alpha}{2 \beta} & 1
\end{pmatrix}
\Comma
\label{eq:vx}
\eeq
\beq
\langle \bm{v}(t) \bm{v}(t) \rangle = \frac{q}{2 \beta}
\begin{pmatrix}
\frac{2 \alpha^2}{\beta} \left [ t -\frac{11}{4 \beta} \right ] + 1
\quad \quad \quad &
\frac{\alpha}{2 \beta} \\
\frac{\alpha}{2 \beta} & 1
\end{pmatrix}
\Teleia
\label{eq:vv}
\eeq
\label{eq:2-point}
\end{subequations}

In a quiescent fluid the fluctuation strength is specified by invoking the
fluctuation-dissipation theorem. Its classical form may be expressed as
(see, for example, Reeks~\cite{mike88}) \beq
\beta \delta_{ij} = \frac{m}{k_B T} \int_0^{\infty} ds \, \langle f_i(s)
f_j(0) \rangle \quad . \label{eq:FDT} \eeq
The use of the classical FDT in sheared systems has been questioned. Whereas
its classical form as shown in Eq.~(\ref{eq:FDT}) has been used in the
past~\cite{sanmiguel,vandeven_jfm80,brady}, it has also been
argued~\cite{colloids} that the fluctuation strength may be
determined by the \textit{ad hoc} requirement that energy equipartition
hold in the local, co-moving reference frame. Energy equipartition in
the local reference frame implies local equilibrium and hence
the Brownian velocity distribution function
becomes locally Maxwellian.
As we show in
Sec.~\ref{sec:eulerian} if the Brownian velocity distribution function is a local
Maxwellian distribution then the Brownian particles behave as an ideal
``particle'' gas with no shear stresses, nor shear viscosity.

More recently, Santamar\'{i}a-Holek et al.~\cite{rubi01} derived a
Fokker-Planck equation for the nonequilibrium distribution function of
Brownian particles in stationary flow. They avoided the use of a Langevin
equation (and the associated problem of the specification of the statistics
of the random force) by using arguments based on mesoscopic nonequilibrium
thermodynamics. They found that the flow modifies the diffusion
tensor in the Fokker-Planck equation
by a term proportional to the imposed velocity gradient.
They concluded that due to this additional term the Fluctuation
Dissipation Theorem in its classical form does not hold for fluctuations
about the nonequilibrium steady state. However, they did not provide an
estimate of the relative magnitude of the nonequilibrium correction to the
classical
FDT~\footnote{The estimate provided in Ref.~\cite{reguera03} refers to the complete
modification of the diffusion coefficient due to the shear under conditions
relevant to nucleation experiments (laminar flow diffusion cloud chamber)
and not to the
relative magnitude of the equilibrium-to-nonequilibrium terms.}.
Such an estimate may be deduced from earlier kinetic theory calculations
of particle motion in a nonuniform light gas. Rodr\'{i}guez et al.~\cite
{dufty83} used Fokker-Planck and Langevin descriptions of fluctuations in
uniform shear flow to conclude that the diffusion tensor in the
Fokker-Planck equation, and the corresponding properties of the random force
in the Langevin description, are modified by a term proportional to the
traceless part of the fluid pressure tensor. Thus, the nonequilibrium
modification of the FDT implies that in a Langevin description the
stochastic properties of the random force are modified by the shear flow.
However, in a Langevin description the time scale of the Brownian
white noise driving force is considered much shorter than the time scale of the
imposed flow suggesting that the nonequilibrium correction would be of the
order of the ratio of the molecular relaxation time to the time scale of the
imposed
shear~\footnote{Time-scale separation is also implicit in the separation of the fluid
velocity into a mean part and a fluctuating part, which gives rise to the
random force in the particle equations of motion Eqs.~(\ref{eq:eom}).}. In
fact, the nonequilibrium correction derived by Rodr\'{i}guez
et al.~\cite{dufty83} can be shown to be proportional to
the ratio of the fluid molecular relaxation time to the time scale of the
imposed shear (a ratio of a microscopic to a macroscopic time scale) by
expressing the (dimensionless) correction in terms of
the velocity gradient (shear rate)
and the ratio of the fluid viscosity to the fluid pressure.
Similarly, Fern\'{a}ndez de la Mora and Mercer~\cite
{delaMorapra82b} used a Chapman-Enskog expansion of the light-gas velocity
distribution function 
to show that the nonequilibrium modification of the Fokker-Planck
equation is of the order of the ratio of the light-gas relaxation
time ($\tau _{f}\thicksim \mu _{f}/p_{f}$) to the macroscopic
fluid deceleration time (the inverse shear rate). Thus, it becomes
of the order of the light-gas Knudsen number and therefore it may
be neglected, only becoming important for rarefied
gases~\cite{delaMorapra82b}. Since in this work we are interested
in cases where time scales are clearly separated, in what follows
we shall use the classical form of the FDT that does not introduce
additional random-force correlations.

The classical form of the FDT and Eq.~(\ref{eq:ff}) specify the fluctuation
amplitude to be
\beq
\frac{q}{2 \beta} = \frac{k_B T}{m} \quad .
\label{eq:v2y}
\eeq
With this identification some of Eqs.~(\ref{eq:2-point}) have been
previously reported~\cite{sanmiguel}, whereas the full time-dependent
correlations for delta-function initial conditions are given in
Ref.~\cite{brady}.
Particle velocity correlations along the $y$ direction (in the long-time
limit), being independent of the shear flow,
satisfy the normal equipartition theorem.
This is expected since the shear flow is only along
the $x$-direction, and the ($x$, $y$) components of the random force have
been assumed uncorrelated. Moreover, a kinetic temperature, as opposed to
the thermodynamic temperature $T$ in Eqs.~(\ref{eq:vv}), may be defined by
relating the average particle kinetic energy to $k_B T_{\text{kin}}$.
Then, the kinetic
temperature along the direction of the shear becomes time-dependent and a
function of particle mass (through the dependence on $\beta$) and of
properties of the fluid flow (the local fluid gradient $\alpha$).

The long-time limit of the equal time correlation $\langle y^2(t) \rangle$
may be used to obtain the Stokes-Einstein expression for the diffusion
coefficient,
\beq
D_0 = \frac{1}{2} \, \frac{d}{dt} \langle y^2(t) \rangle =
\frac{k_B T}{\beta m}
\Teleia
\label{eq:Stokes_Einstein}
\eeq
Note that the diffusion coefficient is first order in $\beta^{-1}$, a result
that will be used later in Sec.~\ref{sec:smoluchowski}.

The shear flow also modifies the time-dependent particle-velocity
autocorrelation functions. Ensemble averages of the formal solutions of the
equations of motions Eqs.~(\ref{eq:solve}) give in the diffusive limit ($t
\gg \beta^{-1}$), neglecting terms involving powers of $\exp(-\beta t)$,
\begin{subequations}
\beq
\langle v_x(t+\tau) v_x(t) \rangle =
\begin{cases}
\frac{k_B T}{m} \left [ e^{-\beta \tau} +
\frac{1}{2} \, \Stk^2 \, \left [ 4 \beta t - 8 -
e^{-\beta \tau} \left ( \beta \tau +3 \right ) \right] \right ]
& \textrm{for} \quad \tau \geq 0 \Comma \\
\frac{k_B T}{m} \left [ e^{\beta \tau} +
\frac{1}{2} \, \Stk^2 \, \left [ 4 \beta \left ( t + \tau \right )
- 8 + e^{\beta \tau} \left ( \beta \tau - 3 \right ) \right] \right ]
& \textrm{for} \quad \tau < 0 \Comma
\end{cases}
\eeq
\beq
\langle v_x(t+\tau) v_y(t) \rangle =
\begin{cases}
2 \, \frac{k_B T}{m} \, \Stk \,
\left [ 1 - \frac{1}{4} e^{-\beta \tau}
\left ( 2 \beta \tau + 3 \right ) \right ]
& \textrm{for} \quad \tau \geq 0 \Comma \\
\frac{k_B T}{m} \, \frac{1}{2} \, \Stk \, e^{\beta \tau}
& \textrm{for} \quad \tau < 0 \Comma
\end{cases}
\eeq
\beq
\langle v_y(t+\tau) v_y(t) \rangle =
\frac{k_B T}{m} e^{-\beta |\tau|}
\quad \forall \tau
\Teleia
\eeq
\end{subequations}
Hence, the combined effects of particle inertia and shear flow
modify both the amplitude of the autocorrelation functions and
their time dependence. The dependence of the autocorrelation
functions on particle inertia and shear rate has been expressed in
terms of the Stokes number, $\Stk = \alpha/\beta$. The shear flow
not only breaks spatial symmetry but also (macroscopic) time
reversibility and stationarity: the particle-velocity
autocorrelation function in the streamwise direction is
nonstationary.
Of course, the velocity correlations perpendicular to the shear decay
exponentially in time as expected for a Gaussian, stationary Markov process,
an Ornstein-Uhlenbeck process. The shear-induced modifications of the
autocorrelation functions become more transparent if the contribution of the
underlying shear flow is subtracted. Specifically, the autocorrelation
functions of the fluctuating particle velocities with respect to the shear
flow, $v_x^{\prime \prime}(t) = v_x(t) -\alpha y(t)$, become (in the
long-time limit)
\begin{subequations}
\beq
\langle v_x^{\prime \prime} (\tau) v_x^{\prime \prime} (0) \rangle =
\frac{k_B T}{m} \, e^{-\beta |\tau|}
\left [ 1 + \frac{1}{2} \, \Stk^2 \, \left (
\beta |\tau | + 1 \right ) \right ]
\quad \forall \tau
\Comma
\eeq
\beq
\langle v_x^{\prime \prime} (\tau) v_y(0) \rangle =
\begin{cases}
- \frac{k_B T}{m} \, \frac{1}{2} \, \Stk \, e^{-\beta \tau}
\left ( 2 \beta \tau + 1 \right )
& \textrm{for} \quad \tau \geq 0  \Comma \\
- \frac{k_B T}{m} \, \frac{1}{2} \, \Stk \, e^{\beta \tau}
& \textrm{for} \quad \tau < 0 \Teleia
\end{cases}
\eeq
\label{eq:stochastic}
\end{subequations}
Thus, the autocorrelation functions of the fluctuating particle velocities
are stationary, the velocity correlation along the shear is symmetric in the
time difference $\tau$, but the cross correlation is non-symmetric in $\tau$%
. Since in the local reference frame time stationarity is recovered the
dependence of the autocorrelation functions on $t$ has been dropped: note,
however, that these expressions are valid only in the long-time limit ($%
\beta t \gg 1$). The time decay of the velocity correlation function along
the flow direction is not a pure exponential, but it is modified by an
algebraic prefactor; hence, the underlying stochastic process $v_x^{\prime
\prime} (t)$ is not an Ornstein-Uhlenbeck process. Furthermore, since the
cross correlation is stationary $\langle v_x^{\prime \prime}(\tau) v_y(0)
\rangle = \langle v_x^{\prime \prime}(0) v_y(- \tau) \rangle$, but since it
is not symmetric $\langle v_x^{\prime \prime}(\tau) v_y(0) \rangle \neq
\langle v_x^{\prime \prime}(0) v_y(\tau) \rangle$. The time asymmetry of the
cross correlation implies that for a negative time difference $\tau < 0$ the
correlation decays exponentially, whereas for $\tau \geq 0$ the correlation
initially increases at short relative times to decrease exponentially at
longer time. The maximum occurs at $\beta \tau = 1/2$.

\begin{figure}[tbp]
\includegraphics[width=16cm,height=12cm]{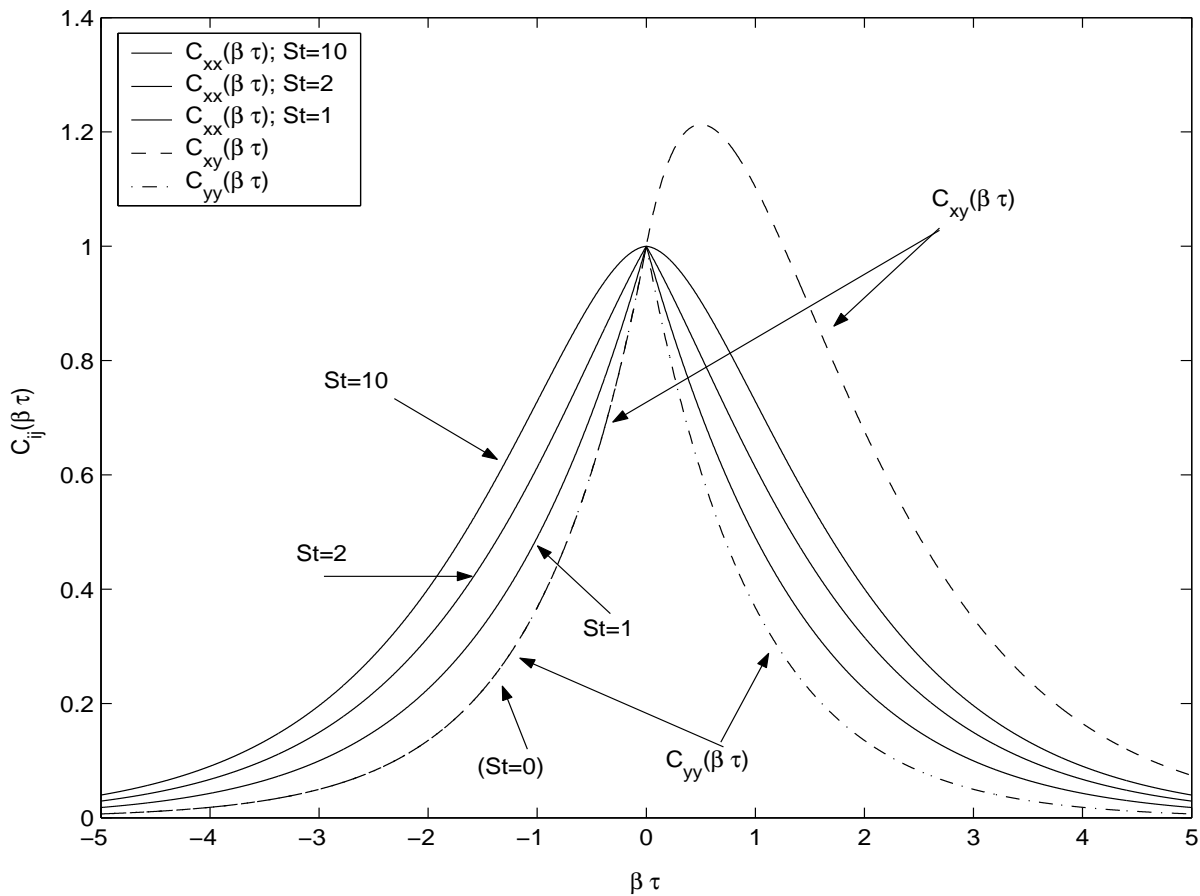}
\caption{Time dependence of the fluctuating particle-velocity
autocorrelation functions in the diffusive limit,
parametrized by the Stokes number: solid lines
are $C_{xx}(\tau)$, the dashed line $C_{xy}(\tau)$, and the dash-dot line is
$C_{yy}(\tau)$.}
\label{fig:fig1}
\end{figure}

These qualitative observations are summarized in Fig.~\ref{fig:fig1}, where
the three correlation functions are compared. The normalized (at the origin)
autocorrelation functions are plotted, $C_{ij}(\tau)= \langle v_i^{\prime
\prime}(\tau) v_j^{\prime \prime}(0) \rangle / \langle v_i^{\prime
\prime}(0) v_j^{\prime \prime}(0) \rangle$ as functions of time rendered
dimensionless by a $\beta$ scaling. The autocorrelation along the
streamwise direction
$C_{xx}(t)$ has been parametrized by three values of Stokes
number; the other two normalized correlations, $C_{xy}(\tau)$ and $%
C_{yy}(\tau)$ are independent of the Stokes number. Note that with
increasing Stokes number the cusp at the origin, characteristic of the
short-time behavior of $C_{yy}(\tau)$, becomes rounded. The $\Stk=10$ curve
is also the value of $C_{xx}(\tau)$ in the limit the $\Stk \rightarrow
\infty$. The velocity autocorrelation functions will be reconsidered in
Sec.~\ref{sec:smoluchowski} since they are related to particle diffusion
coefficients through Green-Kubo relations.

Reeks~\cite{mike_jfm04} in an analysis of particle turbulent dispersion in a
shear flow reports equations similar to Eqs.~(\ref{eq:stochastic}). It can
be shown that the results of Ref.~\cite{mike_jfm04} for the case of
fluid-point dispersion lead to velocity autocorrelation functions identical
to those presented above. Similarly, Eckhardt and Pandit~\cite{eckhardt03} obtained results
formally equivalent to ours for the effect of shear on fluid velocity
fluctuations. Note, however, that for the system under consideration the
effect of the shear flow and particle inertia are described in terms of the
particle Stokes number: hence, for a given shear rate changes of the
Stokes number imply changes of the particle relaxation time (i.e. of the
particle's inertia).

\section{Eulerian description}

\label{sec:eulerian}

The Eulerian description of the motion of $N$, independent and identical
Brownian particles in terms of the average phase space density
$P(\xvec, \vp ; t)$ is an alternative to
the Lagrangian description in terms of stochastic
differential equations. The Fokker-Planck equation associated with the
particle equations of motion Eqs.~(\ref{eq:eom}) may be derived in numerous
ways, including the elegant functional derivation of Refs.~\cite
{keith_thesis,keith99}. It becomes
\beq
\frac{\partial}{\partial t} P + \frac{\partial}{\partial \xvec} \cdot
\left
( \vp P \right ) - \beta \, \frac{\partial}{\partial \vp} \cdot
\left [ \left ( \vp - \uf \right ) P \right] = \beta \, \frac{k_B T}{m} \,
\frac{\partial^2}{\partial \vp^2} P
\quad ,
\label{eq:generalfpe}
\eeq
where the fluctuation amplitude, defined in Eq.~(\ref{eq:v2y}), was used.
The probability density function $P$ is normalized on the total number of
particles. As mentioned before, Eq.~(\ref{eq:generalfpe}) differs from the
Fokker-Planck equation derived by Santamar\'ia-Holek et al.~\cite{rubi01} in
the diffusive term since we used the classical FDT.

The probability density function $P$ is used to define average
quantities. In particular, the mean, local particle velocity,
a quantity defined as in kinetic theory of gases~\cite{resibois} is
\beq
\rho (\xvec;t) \thinspace \overline{\bm{v}} (\xvec; t) = m \thinspace \int d \vp
\thinspace P(\xvec, \vp; t) \thinspace \bm{v} \triangleq m \langle n v
\rangle
\quad ,
\eeq
with $n$ the ``instantaneous'' particle number density
and $\rho (\xvec;t)$ the local, average
particle mass density,
\beq
\rho (\xvec; t) = m \thinspace \int d \vp \thinspace P(\xvec, \vp; t)
\label{eq:density_def}
\quad .
\eeq
The fluctuating component of the particle velocity field,
defined with respect to the mean particle
velocity~\footnote{Primed variables refer to
fluctuating velocities with respect to the mean
particle velocity, whereas double-primed variables are defined with respect
to the carrier fluid velocity, see Eqs.~(\ref{eq:stochastic}).},
will be denoted by $\vprime = \bm{v} -\overline{\bm{v}}$.

These local averages, denoted by an overbar and known as density-weighted
averages in two-fluid descriptions of dispersed flows, are averages with
respect to a local, properly normalized velocity probability density
distribution
\beq
\psi(\xvec, \vp; t) =\frac{P(\xvec, \vp; t)}{\int d \vp \, P(\xvec, \vp; t)}
\quad .
\label{eq:psi}
\eeq
The integral of $\psi$ over particle velocities is, thus, unity.

The appropriate equations to describe particle dispersion, also known as
``continuum'' or ``hydrodynamic'' equations, are obtained by taking
velocity moments of the Fokker-Planck equation.
In particular, the particle mass conservation equation
is obtained by multiplying
Eq.~(\ref{eq:generalfpe}) by $m$, integrating over particle velocities and
using Eq.~(\ref{eq:density_def}). Multiplying the
Fokker-Planck equation by $m\vprime$ and realizing that
$m\,\langle n\vprime \vprime \rangle =\rho \overline{\vprime \vprime}$
[see the definition Eq.~(\ref{eq:psi})] gives the
momentum conservation equation. Accordingly,
they become (see also~\cite{keith_thesis})
\begin{subequations}
\beq
\frac{d \rho}{d t} = - \rho \boldnabla \cdot \vmeanb
\Comma
\label{eq:zeroth}
\eeq
\beq
\frac{d \overline{\bm{v}}}{dt} =
\beta ( \uf - \vmeanb ) -
\frac{1}{\rho} \frac{\partial}{\partial \bm{x}} \cdot
\left ( \rho \overline{\vprime \vprime} \right )
\Comma
\label{eq:init_momentum}
\eeq
\label{eq:continuum}
\end{subequations}
where the total derivative
$d/dt=\partial /\partial t+\overline{\vp} \cdot \boldnabla$
describes changes with respect to the mean particle velocity.
Equations~(\ref{eq:continuum}) are a special case for a white noise random
force of the general continuum equations for the dispersed phase in
inhomogeneous flows derived by Reeks~\cite{mike92}.

The momentum equation Eq.~(\ref{eq:init_momentum}) may be compared to the
momentum equation that expresses momentum balance in the
particulate phase~\cite{delaMorapra82a,konstandopoulos90}
\beq
\frac{d \overline{\bm{v}}}{dt} = \beta (\uf -\vmeanb) - \frac{1}{\rho}
\boldnabla \cdot \tensor{\bm{P}}_p
\quad ,
\label{eq:momentum}
\eeq
where $\tensor{\bm{P}}_p$ is the unspecified, particle-phase total pressure tensor.
Comparison yields $\tensor{\bm{P}}_p= \rho \overline{\vprime \vprime}$.
It is easy to show [via Eq.~(\ref{eq:psi})] that the expression for the
particle-phase pressure tensor is identical to the pressure tensor as
defined in kinetic theory~\cite{resibois} or in mesoscopic nonequilibrium
thermodynamics~\cite{rubi01}
\beq
\tensor{\bm{P}}_p= m \, \int d \vp (\vp -\overline{\vp}) (\vp -\overline{\vp})
\, P(\xvec, \vp; t)
\quad .
\eeq

It is apparent from the previous discussion that the continuum equations
and the identification of the particle-phase pressure tensor with the
particle covariances only require that the random force be white in time.
Thus, the previous expressions are valid for a general flow field and
not only for a linear flow field (as long as the random force is white).
In what follows we restrict
the calculation to a linear flow field, eventually
evaluating average quantities (for example, the mean particle velocity
and the particle-phase pressure tensor) for the specific
case of a simple shear flow, as discussed and justified in the Introduction.

For a linear flow field the Fokker-Planck equation defines a
Gaussian process for $\left[ \xvec,\vp\right]$ and thus it has an
analytic solution, see, for
example, Refs.~\cite{keith_thesis,swailes97}. The Gaussian solution may be used
to evaluate explicitly the density-weighted ensemble averages;
Swailes and Darbyshire~\cite{swailes97} report the analytic
expressions. For Brownian particles injected at the origin of the
co-ordinate system with zero initial velocity the
spatially-dependent particle concentration is
(Ref.~\cite{swailes97} presents the analytic solutions for
arbitrary initial conditions)
\beq
\rho(\xvec; t) = \frac{1}{2\pi
\left [ \det \left ( \langle \xvec \xvec \rangle \right ) \right
]^{1/2}} \exp \left [ -\frac{1}{2} \xvec^{\textrm{T}} \cdot \langle \xvec \xvec
\rangle^{-1} \cdot \xvec \right ] \quad ,
\label{eq:rho}
\eeq
the mean particle velocity
\beq
\overline{\bm{v}} (\xvec; t) =
\langle \vp \xvec \rangle \cdot \langle \xvec \xvec \rangle^{-1} \cdot
\xvec \quad ,
\label{eq:vmean}
\eeq
and the particle-velocity covariances
\beq
\overline{\vprime \vprime} (t) = \langle \vp \vp
\rangle - \langle \vp \xvec \rangle \cdot \langle \xvec \xvec
\rangle^{-1} \cdot \langle \xvec \vp \rangle \quad ,
\label{eq:vprimevprime}
\eeq
where the superscript in Eq.~(\ref{eq:rho}) denotes
transpose.  Thus, for the Gaussian process the
mean particle velocity is linear in $\xvec$, and the particle
covariances are spatially independent. Analytic expressions for
the long-time behavior of the mean-particle velocity and its
approach to the steady-state value (the carrier flow velocity) are
presented in the
Appendix. 
For completeness note that explicit evaluation of the necessary correlation
functions in the long-time limit shows that $\lim_{t \rightarrow \infty} \,
\overline{\vprime \vprime} = \langle (\vp- \uf)^2 \rangle$ since $\lim_{t
\rightarrow \infty}\langle \vp \rangle = \uf$.

The long-time particle-phase pressure tensor may then be evaluated
using Eqs.~(\ref{eq:2-point}) and (\ref{eq:vprimevprime}):
expressed as a function of the Stokes number, it becomes
\beq
\tensor{\bm{P}}_p = \rho \frac{k_B T}{m} \left [ \left ( 1+
\frac{\Stk^2}{4}\right ) \tensor{\bm{I}} - \frac{\Stk}{2}
\begin{pmatrix}
0 &  1 \\
 1 & 0
\end{pmatrix}
+ \frac{\Stk^2}{4} \begin{pmatrix}
1 & 0 \\
0 & -1
\end{pmatrix}
\right ] \quad .
\label{eq:Pi}
\eeq
The particle pressure tensor has been decomposed into a form reminiscent of
the pressure tensor of a simple liquid undergoing laminar plane Couette
flow~\cite{hess84}. The first term is isotropic and proportional to the identity
tensor; the other two terms constitute what has been called~\cite{hess83}
the ``friction pressure tensor''~\footnote{If the momentum equation
had been expressed in terms of the total
particle-phase stress tensor, the negative of $\tensor{\bm{P}}_p$ as defined
by Eq.~(\ref{eq:momentum}), then these terms would constitute the deviatoric
particle stress tensor.}. The first part of the isotropic term gives the
ideal gas pressure of a collection of Brownian particles (``Brownian
particle gas''), a result that has been previously postulated on
phenomenological arguments~\cite{ramshaw79,delaMora82}. The ideal gas
pressure is modified by a correction dependent on the Stokes number. The
other terms, usually neglected, arise from the particle viscous stresses.
The second term is proportional to the symmetric rate-of-strain fluid
tensor, $2 \, \tensor{\bm{\epsilon}} = \vec{\nabla} \vec{u} +
\left ( \vec{\nabla} \vec{u} \right )^{\textrm{T}}$. Thus, the
(negative) proportionality constant
defines the shear viscosity of the particle phase to be
\beq
\eta_p^{\textrm{shear}} =\frac{1}{2} D_0 \rho
\quad .
\label{eq:etaB}
\eeq
Hence, the Brownian shear viscosity, being independent of the shear rate, is
a conventional Newtonian viscosity. An alternative expression for
Eq.~(\ref{eq:etaB}) is that the particle Schmidt number, the ratio of
momentum diffusivity to mass diffusivity, is
$Sc_p= (\eta_p^{\text{shear}}/\rho)/D_0=1/2$,
as is the case for particle motion in a turbulent flow in
which the particle response time is much larger than the turbulent time
scale~\cite{mike_jfm04}. The third term, which is absent in Newtonian
fluids, shows that the particle phase exhibits non-Newtonian rheology, as
has been remarked for sheared simple liquids~\cite{hess84,hess83}. Note that
the last term, as well as the correction to the ideal gas pressure, is
second order in the Stokes number. Hence to leading order in $\Stk$ the
viscous part of the particle pressure tensor is traceless, and the particle
phase behaves as a Newtonian fluid. Non-Newtonian behavior becomes evident
only to second order. Moreover, according to Eq.~(\ref{eq:Pi}) [and Eq.~(\ref
{eq:Stokes_Einstein})] in the absence of diffusion, for
sufficiently massive particles, $\tensor{\bm{P}}_p=0$.

As remarked in Sec.~\ref{sec:lagrangian} if local equilibrium is
assumed, as for example in Ref.~\cite{colloids}, $\langle (\bm{v}-\uf)(\bm{v} - \uf)
\rangle \triangleq k_B T \tensor{\bm{1}} /m $, the Brownian
particle distribution becomes locally Maxwellian. Then,
from Eqs.~(\ref{eq:vv}) the fluctuation strengths can be calculated
to form a symmetric tensor dependent on the shear rate.
As extensively argued in Sec.~\ref{sec:lagrangian} the dependence
of the fluctuation strengths on the shear rate
implies that the externally imposed shear modifies
the properties of the random force, a condition
that is unlikely to hold when the relevant time scales
(inverse shear rate and molecular relaxation time scale)
are clearly separated.
Moreover, for a local equilibrium assumption
a similar calculation shows that the
particle-phase pressure tensor (in the
long-time limit) contracts to the
ideal gas result,
$\tensor{\bm{P}}_p^{\text{l.eq.}} = \rho (\xvec; t) k_B T \tensor{\bm{1}}/m$,
i.e. there is no viscous shearing.

Equation~(\ref{eq:Pi}) along with Eqs.~(\ref{eq:continuum})
are our main result. They
are analytic, non-perturbative expressions that describe the long-time
behavior of the dispersed phase: they incorporate the effects of the flow
field, and the combined effects of particle diffusion and particle inertia
on the particle-phase total pressure tensor. However, the momentum equation
is coupled to the mass conservation equation. The two (steady-state)
equations are, usually, decoupled by performing a low Stokes-number
expansion of the momentum equation to obtain a perturbative expression for
the mean particle velocity field in terms of the fluid velocity and its
gradients~\cite{delaMora82}. Substitution of the resulting mean particle
velocity into the particle mass conservation equation then gives the
associated convective-diffusion equation~\cite{ramshaw79}.
For the case of a simple shear this procedure leads
to a generalized Smoluchowski equation [since the
local average density is defined as an integral of the
average phase space density $P$, Eq.~(\ref{eq:density_def})],
as we show in the following Section.

\section{Smoluchowski equation}

\label{sec:smoluchowski}

The momentum equation~(\ref{eq:init_momentum}), an equation valid
for a general flow field and a white random force, may be
re-arranged to obtain an explicit expression for the total time
derivative of the mean particle-velocity field (also referred to
as the inertial acceleration term). For the specific case of a
simple shear differentiation of the analytic, time-dependent
expression for the particle concentration Eq.~(\ref{eq:rho}) leads
to
\beq
\rho \xvec = - \langle \xvec \xvec \rangle \cdot
\frac{\partial \rho}{\partial \xvec} \quad ,
\label{eq:linearflow2}
\eeq
which substituted into the equation
for the mean particle velocity Eq.~(\ref{eq:vmean}) expresses the
particle mean velocity in terms of the density gradient
\beq
\overline{\bm{v}} = - \langle \vp \xvec \rangle \cdot
\frac{1}{\rho} \frac{\partial \rho}{\partial \xvec} \quad .
\label{eq:vmean_gradient}
\eeq
Moreover, a variant of
Eq.~(\ref{eq:linearflow2}), obtained by multiplying it by the
shear-rate tensor $\tensor{\bm{\alpha}}$, determines the carrier
flow velocity
\beq
\rho \uf = - \langle \uf \xvec \rangle \cdot
\frac{\partial \rho}{\partial \xvec} \quad .
\label{eq:rhouf}
\eeq
Equations~(\ref{eq:vmean_gradient}) and (\ref{eq:rhouf}), derived
from the time-dependent solution for the local particle density,
are generally valid for a simple shear and not only in the
long-time limit. Their substitution into the momentum equation,
Eq.~(\ref{eq:init_momentum}), along with the realization that for
a linear flow field (and, in particular, a simple shear) the
velocity covariances are spatially independent, leads to
\beq
\rho \frac{d \overline{\bm{v}}}{dt} = \left [ \beta \langle (\bm{v}
-\uf) \xvec \rangle -\overline{\vprime \vprime} \right ] \cdot
\frac{\partial \rho} {\partial \xvec} \triangleq \tensor{\bm{M}}
\cdot \frac{\partial \rho}{\partial \xvec} \quad .
\label{eq:defineM}
\eeq
Hence, the inertial acceleration term in a
simple shear may be explicitly evaluated: specifically,
Eqs.~(\ref{eq:2-point}) along with Eq.~(\ref{eq:v2y}) and
Eq.~(\ref{eq:vprimevprime}) provide analytic expressions for the
necessary correlations (in the diffusive limit). Their substitution
into Eq.~(\ref{eq:defineM}) yields
\beq
\tensor{\bm{M}} = \frac{k_B T}{m} \Stk
\begin{pmatrix}
-2 \Stk & -1 \\
1 & 0
\end{pmatrix}
\quad .
\label{eq:matrixM}
\eeq
Thus, the proportionality matrix $\tensor{\bm{M}}$ is of the
order of the Stokes number; note, however, that for a simple shear
the convective derivatives of the flow
field [$(\uf \cdot \nabla) \uf$] vanish.
In this respect the simple shear differs from
the symmetric and antisymmetric shear, the other
possible representations of a general two-dimensional
linear flow field.
Equation~(\ref{eq:matrixM}) for the inertial acceleration term
and Eq.~(\ref{eq:Pi}) for the particle-phase pressure tensor
were derived for the specific case of a simple shear. Hence,
they may be substituted into the general, particulate-phase,
momentum-conservation equation Eq.~(\ref{eq:momentum}),
or equivalently into Eq.~(\ref{eq:init_momentum}), rewritten as
\beq
\vmeanb = \uf
- \frac{1}{\beta \rho} \boldnabla \cdot \tensor{\bm{P}}_p - \frac{1}{\beta}
\, \frac{d \overline{\bm{v}}}{dt}
\quad ,
\label{eq:vmean_general}
\eeq
to obtain the analytic, non-perturbative expression for
the mean particle velocity field in the diffusive limit ($t \gg \beta^{-1}$)
\begin{align}
\overline{\bm{v}} & = \uf - \langle (\bm{v} -\uf) \xvec \rangle
\cdot \frac{1}{\rho} \frac{\partial \rho}{\partial \xvec}
\label{eq:low_Stk} \\
& \overset{\beta t \gg 1}{=} \uf - D_0 \left [ \tensor{\bm{1}} - \frac{3}{2} \Stk \begin{pmatrix}
\Stk & 1 \\ - \frac{1}{3} & 0 \end{pmatrix} \right ] \cdot \frac{1}{\rho}
\frac{\partial \rho}{\partial \xvec}
\quad .
\label{eq:final_lowStk}
\end{align}
Note that Eq.~(\ref{eq:low_Stk}) may also be derived by a
judicious combination of Eqs.~(\ref{eq:vmean_gradient}) and
(\ref{eq:rhouf}) as can be seen by expanding the ensemble average
$\langle (\bm{v} -\uf) \xvec \rangle$. The  alternative derivation
presented earlier, however, is more instructive because it allows
the identification of the contribution of the particle-phase
pressure tensor and the inertial acceleration to the diffusion
coefficient, as discussed in what follows.

To leading order in the particle relaxation time ($\beta^{-1}$) the mean
particle velocity is the fluid velocity modified by a diffusion term (Fick
diffusion). Fick diffusion stems from the ideal part
of the particle-phase pressure tensor. The correction to Fick's law arises from
the inertial acceleration and the particle viscous stresses. In fact, the contribution
of the inertial acceleration term in Eq.~(\ref{eq:vmean_general}) is of the
same order as the contribution from the particle-phase viscous stresses.
Either contribution to the mean-particle velocity is at least
second order in $\beta^{-1}$ [see, also,
Eq.~(\ref{eq:Stokes_Einstein})].
Hence, in a simple shear it is inconsistent to retain particle viscous
stresses and neglect inertial acceleration, and vice versa.

The convective-diffusion equation (Smoluchowski equation) associated with
the two coupled Eqs.~(\ref{eq:continuum}) is obtained
by substituting the mean particle velocity Eq.~(\ref{eq:low_Stk})
into the mass conservation equation to obtain
\beq
\frac{\partial
\rho}{\partial t} + \boldnabla \cdot (\rho \uf) = \langle (\bm{v}
-\uf) \xvec \rangle \cdot \frac{\partial^2 \rho}{\partial \xvec^2}
\quad .
\label{eq:conv_diff}
\eeq
Equation~(\ref{eq:conv_diff})
defines the diffusion tensor
$\tensor{\bm{D}}= \langle \bm{v}^{\prime \prime} \xvec \rangle$
that depends on both
particle inertia and the shear rate through the Stokes number. In
the long-time limit it is given by the second term of the RHS of
Eq.~(\ref {eq:final_lowStk}). Note that the diffusion tensor, as
well as the matrix expression for the inertial acceleration term
Eq.~(\ref{eq:matrixM}), is not symmetric, reflecting the
symmetry-breaking effect of the imposed shear. Therefore, the
long-time generalized Smoluchowski equation becomes \beq
\frac{\partial \rho}{\partial t} + \alpha y \frac{\partial
\rho}{\partial x} = D_0 \left (1 -\frac{3}{2} \Stk^2 \right )
\frac{\partial^2 \rho}{\partial x^2} - D_0 \, \Stk
\frac{\partial^2 \rho}{\partial x \partial y} + D_0
\frac{\partial^2 \rho}{\partial y^2} \quad .
\label{eq:Smoluchowski} \eeq Since the inertial acceleration term
is anti-symmetric it only modifies the diffusion coefficient in
the streamwise direction; the modification is second order in the
Stokes number. The first order correction through the appearance
of the cross derivative term arises solely from the viscous part
of the particle pressure tensor. The same first order correction
to the Smoluchowski equation was derived by Subramanian and
Brady~\cite{brady}, who, however, did not calculate higher order
corrections.

\begin{figure}[tbp]
\includegraphics[width=14cm,height=10cm]{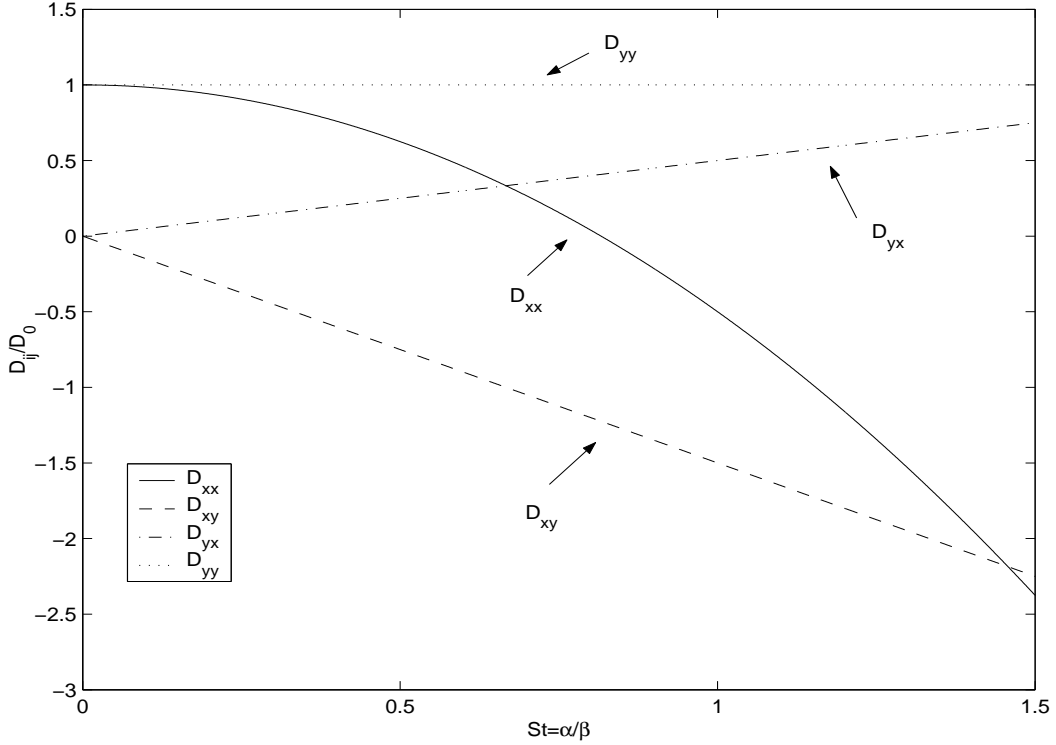}
\caption{Dependence of the long-time diffusion coefficients on particle inertia
expressed in terms of the Stokes number. The components $D_{yx}$ (dash-dot
line) and $D_{yy}$ (dotted) are always positive, $D_{xy}$ (dashed) always
negative, whereas the sign of $D_{xx}$ (solid) changes as a function of the
Stokes number.}
\label{fig:fig2}
\end{figure}

The sign of the long-time diffusion coefficient along the flow
direction, $D_{xx}$ in Eq.~(\ref{eq:Smoluchowski}),
depends on the value of the shear rate. For large shear rates it
becomes negative, the critical Stokes number being $\Stk=\sqrt{3/2}
$. The long-time limit of the off-diagonal diffusion coefficients is such
that $D_{yx}$ is always positive, whereas $D_{xy}$ is always negative [See
Eq.~(\ref{eq:final_lowStk})]. The other diffusion coefficient, $D_{yy}$, is
always positive. The dependence of the diffusion coefficient on particle
inertia (via the Stokes number) is shown in Fig.~\ref{fig:fig2}.

The existence of negative diffusion coefficients does not violate any
stability conditions since the coefficients that appear in the
convective-diffusion equation refer to
the particle diffusive flux with respect to the
underlying carrier flow. As expected from general stability arguments the
total diffusive flux defined with respect to a fixed reference frame, and
not with respect to the carrier flow, is always positive. The mean particle
velocity expressed in terms of the density gradient,
Eq.~(\ref{eq:vmean_gradient}), defines the total diffusion coefficients,
that in the diffusive, long-time limit become
\beq
\tensor{\bm{D}}_{\textrm{tot}} \triangleq \langle \vp \xvec \rangle = D_0
\begin{pmatrix}
1+\alpha^2 t^2 & 2 \alpha t \\
\frac{\alpha}{\beta} & 1 \\
\end{pmatrix}
\quad .
\label{eq:Dtotal}
\eeq
Therefore, the total diffusion tensor is time dependent, non-symmetric,
the corresponding matrix positive definite,
and its components are always positive,
thereby assuring the stability of the system. The
shear-induced modification of the total diffusion coefficients, in
particular the quadratic time dependence of the
streamwise diffusion coefficient, has been noted before, for
example in Refs.~\cite{hess84,sanmiguel,vandeven_jfm80}.
Note that the symmetric part of $\tensor{\bm{D}}_{\textrm{tot}}$
gives the total diffusion coefficients as determined from
the rate of change of the particle mean square displacement.

The long-time diffusion coefficients may also be obtained from the velocity
autocorrelation functions presented in Sec.~\ref{sec:lagrangian}. The
following Green-Kubo relations, also known as Taylor's diffusion formulae
for diffusion via continuous movements in theories of turbulent particle
dispersion, express the diffusion coefficients in terms of the velocity
autocorrelation functions
\beq
\tensor{\bm{D}}(t) = \langle \bm{v}^{\prime \prime}(t) \xvec(t) \rangle =
\int_0^t \, dt_1 \, \langle \bm{v}^{\prime \prime}(t) \bm{v}(t_1) \rangle
\Teleia
\label{eq:Dtime}
\eeq
In the diffusive limit they become
\beq
\tensor{\bm{D}} = \int_0^{\infty} \, dt_1 \, \lim_{t \rightarrow \infty} \,
\left [ \langle \bm{v}^{\prime \prime}(t) \bm{v}(t-t_1) \rangle \right ]
\Comma
\label{eq:Dinfty}
\eeq
expressions that reproduce the diffusion tensor Eq.~(\ref{eq:final_lowStk})
[see, also, Eq.~(\ref{eq:DxxGK})].

Equation~(\ref{eq:Dinfty}), in addition to providing the long-time diffusion
coefficients in terms of velocity autocorrelations, allows a physical
interpretation of the negative diffusion coefficients. Specifically, the
long-time limit of $D_{xy}$ is always negative because a particle crossing
the mean shear flow with positive $v_y$ will have negative streamwise
fluctuations (as the mean flow increases with $y$). Along the streamwise
direction $D_{xx}$ becomes negative for large shear rates because streamwise
fluctuations $v_x^{\prime \prime} = v_x - \alpha y$ become negative for
large values of the shear rate $\alpha$ (for $v_x > 0$).

The results presented in this Section may be critically compared to
earlier analyses of diffusive motion in nonuniform flows. We showed,
via the derivation of analytic, long-time expressions
that in a simple shear the inertial acceleration term
(total derivative of the mean particle velocity) and the
particle-phase viscous stresses have to be treated consistently.
In particular, we argued that in perturbative evaluations of, for
example, the Brownian particle diffusive flux in a simple shear,
it is inconsistent to retain the particle-phase viscous stresses
and neglect inertial acceleration, and vice versa.
Previous studies  differ from ours in the way these two
terms ($\tensor{\bm{P}}_p$ and $d \overline{\bm{v}}/dt$)
are treated (and to a lesser degree in the choice of the
underlying flow field).

Specifically, Fern\'{a}ndez de la Mora and Rosner~\cite{delaMora82}
considered the effect of inertia on diffusional deposition
for a general flow field. They
solved the steady-state momentum equation, Eq.~(\ref{eq:vmean_general}),
to leading order in $\beta^{-1}$ keeping only
the ideal-gas part of $\tensor{\bm{P}}_p$ and the leading order
contribution from the inertial acceleration term [$\beta^{-1} \, (\uf \cdot \nabla) \uf$].
For a general flow field this is a consistent approximation: higher order
corrections would require an expression for the particle viscous
stresses. For a simple shear their result reduces to ours to the
same order [$\overline{\bm{v}} = \uf - D_0 \nabla \log \rho +
O(\beta^{-2})$] since, as noted earlier, the convective
derivatives of the flow field vanish in a simple shear.

Ramshaw~\cite{ramshaw79} also analyzed Brownian motion
in a general flow field, but his analysis
differs from ours in that particles of negligible
inertia were considered. He developed a leading order in $\beta^{-1}$ expansion
neglecting both viscous stresses and inertial acceleration,
but keeping other forces, e.g. thermophoresis.
Ramshaw~\cite{ramshaw81} extended the original derivation
to include phenomenologically viscous stresses in the particle and
mixture (particle and fluid) momentum equations
retaining however the assumption of negligible particle inertia.
He found that inclusion of viscous stresses in the mixture
equation, and evaluation of the suspension viscosity by Einstein's
formula, leads to an additional term in the diffusion tensor
proportional to the particle-phase volume fraction. Our calculation
considers an infinitely dilute suspension, and hence this
additional term is absent from Eq.~(\ref{eq:final_lowStk}).
Moreover, his calculation shows that inclusion of viscous stresses
only in the particle-phase momentum equation does not modify the diffusion
coefficient if the particle shear viscosity is evaluated as if the
particles were an ideal gas (and, hence, the shear viscosity
is independent of the particle mass density).

The analysis of Santamar\'ia-Holek et al.~\cite{rubi01}, who
considered diffusion in a simple shear as in this work, differs
from ours primarily in the use of the nonequilibrium FDT (as
summarized in Sec.~\ref{sec:lagrangian}). Moreover, the diffusion
tensor was evaluated perturbatively in the particle relaxation
time ($\beta^{-1}$) retaining the viscous part of the
particle-phase pressure tensor and neglecting the contribution
from the inertial acceleration term. As shown earlier, when the
classical form of the FDT is used these two terms contribute to
the same order ($\beta^{-2}$) in the perturbation expansion.
However, the resulting, leading order correction to the
Smoluchowski equation is identical to ours (up to the
nonequilibrium modification) because the inertial acceleration
(being anti-symmetric) contributes only to the highest order. A
similar remark applies to the leading order correction to the
Smoluchowski equation derived by Subramanian and
Brady~\cite{brady} (who neglected the nonequilibrium modification,
as in this work).

\section{Conclusions}

\label{sec:conclusions}

The primary question addressed in this work is how particle inertia modifies
diffusional transport of particles in a nonuniform flow.
The usual convective-diffusion equation describes diffusional transport in
the limit where the particles follow the fluid streamlines. It is an
Eulerian continuum equation that provides a computationally efficient method
to calculate particle transport (and deposition); however inertial effects
are neglected. Inertial transport, where particle trajectories
deviate considerably from the fluid streamlines, is best calculated via a
Lagrangian description in terms of the particle equations of motion.
Diffusive particle motion may be incorporated in the Lagrangian formulation
through the addition of a random force, but the numerical solution of the
resulting stochastic differential equations is computationally intensive. It
is, thus, desirable to derive a continuum equation valid in
the transition regime between the diffusion limit and the inertia-dominated
limit, incorporating both particle-transport mechanisms.


We considered the coupled diffusive and inertial motion of non-interacting
Brownian particles in a simple inhomogeneous fluid flow, a simple shear
flow. The long-time, diffusive, behavior of the system was investigated
neglecting the short-time regime. Even though the choice of a
simple shear as the underlying carrier flow is restrictive
[for example, the convective derivatives of the
flow vanish, $( \uf \cdot \nabla) \uf = 0$, a condition that
does not hold for a symmetric or antisymmetric shear]
the choice was motivated by numerous
previous investigations of Brownian motion in  such a flow.
More importantly, the Fokker-Planck equation
associated with the particle equations of motion
in a linear flow field is of the linear type~\cite
{dufty83}, and hence analytically solvable. The solution of the
Fokker-Planck equation was used to obtain analytic expressions for average
particle properties, for example the mean particle velocity and
particle-velocity correlations. These analytic solutions depend on
equal-time ensemble averages that were determined from the formal solutions
of the Langevin equations for particle motion. The classical form of the
Fluctuation Dissipation Theorem was used to determine the strength of the
random force correlations, neglecting corrections of the order of the ratio
of the fluid molecular relaxation time to the time scale of the imposed shear.
We showed that the long-time, time-dependent particle-velocity
autocorrelation along the streamwise direction is nonstationary. The fluctuating
velocity autocorrelations were determined to be stationary in time, but the
cross correlation was non-symmetric in the time difference, reflecting the
combined effect of particle inertia and shear on particle-velocity
fluctuations.

We used the analytic solution of the Fokker-Planck equation, in conjunction
with its first two velocity moment equations, to obtain [in the
diffusive limit ($t \gg \beta^{-1}$)] the generalized
convective-diffusion equation (generalized
Smoluchowski equation) that incorporates inertial effects on diffusional
transport for dilute suspensions. The coupling of particle inertia to the
fluid flow introduces a shear-dependent, linear in the (particle) Stokes number, cross
derivative term, and an additional term
along the streamwise direction, quadratic in the Stokes number.
The associated diffusion tensor, thus, depends
on the shear rate and particle inertia; the
diffusion coefficient along the streamwise
direction flow was found to become negative for large particle relaxation times (or,
equivalently, for large shear rates), whereas one of the cross diffusion
coefficient was determined to be always negative, the other two being always
positive. We argued that stability conditions are not violated since the
total diffusion coefficients 
(not those with respect to the carrier flow)
that measure the rate of
change of particle mean square displacement
were determined to be always positive.

We showed that in a simple shear, and in the long-time limit, the contribution
of the inertial acceleration term to the diffusion tensor in the
generalized Smoluchowski equation is of the same order in the Stokes number
as that of the viscous part of the
particle-phase pressure tensor. Thus, in perturbative evaluations of, for
example, the Brownian particle diffusive flux, it is inconsistent to retain
the particle-phase viscous stresses
and neglect the inertial acceleration term, and vice versa.

As part of the derivation of the Smoluchowski equation we calculated the
particle-phase total pressure tensor that was determined to be second order
in the Stokes number. Similar to the pressure tensor of simple sheared
liquids, the pressure tensor was decomposed into three parts: a part
proportional to the identity tensor that gives the ideal pressure of a gas
of Brownian particle with an additional term due to particle inertia, and
two terms that arise from the particle viscous stresses. We found that the
particle phase has a conventional shear viscosity, but it behaves as a
non-Newtonian fluid if second order effects in the Stokes number are
considered.

These results will be extended to other two-dimensional linear flows
in future work.

\section{Acknowledgements}

Y.D. acknowledges partial financial support from the European Commission
through project URBAN AEROSOL under contract No. EVK4-CT-2000-00018. M.W.R.'s
work was supported by a Marie Curie Fellowship of the European Community
Programme Improving Human Research Potential under contract No.
HPMF-CT-2002-02051.

\appendix*

\section{Approach to the steady state}

The time-dependent, equal-time correlation functions in conjunction with
Eq.~(\ref{eq:vmean}) [or Eq.~(\ref{eq:vmean_gradient})] may be used to
calculate the approach of the mean particle velocity to its steady-state
value. The required calculations were performed
with \textit{Mathematica\/}~\cite{mathematica}. Accordingly,
the mean particle velocity approaches the
carrier fluid velocity as follows
\begin{subequations}
\begin{align}
\lim_{t \rightarrow \infty} \overline{v}_x & =
\alpha y \left ( 1 -\frac{3}{\beta t} \right )
+ \left [ \frac{9 x}{2 \beta} - \frac{3 y}{\alpha}
\left ( 1 - \frac{3}{2} \Stk^2
\right ) \right ] \, t^{-2} \,
+ O \left ( t^{-3} \right )
\Comma \\
\lim_{t \rightarrow \infty} \overline{v}_y & =
\frac{2 y}{t} - \frac{3}{\alpha}
\left ( x + \frac{\Stk}{2} y \right )
\, t^{-2} \,
+ O \left ( t^{-3} \right )
\Teleia
\end{align}
\label{eq:vmean_ness}
\end{subequations}
Note that the long-time limit and the vanishing shear-rate limit do not
commute since
\begin{subequations}
\begin{align}
\lim_{t \rightarrow \infty} \left ( \lim_{\alpha \rightarrow 0} \overline{v}_x \right ) & =
\frac{x}{2 t} \left ( 1 + \frac{3}{2 \beta t} \right )
+ \frac{3 y}{4} \left [ 1 -\frac{2}{3 \beta t}
- \frac{19}{12 (\beta t)^2} \right ] \, \alpha
+ O(\alpha^2)
\Comma \\
\lim_{t \rightarrow \infty} \left ( \lim_{\alpha \rightarrow 0} \overline{v}_y \right ) & =
\frac{y}{2 t} \left ( 1 + \frac{3}{\beta t} \right )
- \frac{x}{4} \left [ 1 -\frac{2}{\beta t}
- \frac{5}{4 (\beta t)^2} \right ] \, \alpha
+ O(\alpha^2)
\Teleia
\end{align}
\label{eq:alphalimit}
\end{subequations}
The zeroth order terms of Eqs.~(\ref{eq:alphalimit}) give the mean particle
velocity of Brownian particles in a quiescent fluid.

The time-dependent approach of the mean particle velocity to its
steady-state value may be used to estimate the relative importance of the
two terms in the total derivative of the mean-particle velocity,
namely $\partial /\partial t$ and $(\vmeanb \cdot \boldnabla)$. Appropriate
differentiation of Eqs.~(\ref{eq:vmean_ness}), along with
\begin{subequations}
\beq
\frac{\partial \log \rho}{\partial x} =
\frac{1}{\Stk} \, \frac{m}{k_B T} \, \frac{3 y}{t^2} + O(t^{-3})
\Comma
\eeq
\beq
\frac{\partial \log \rho}{\partial y} =
- \frac{2 y}{D_0 t} + O(t^{-2})
\Comma
\eeq
\end{subequations}
shows that the explicit time derivative vanishes in the long-time limit (to
leading order in $t^{-1}$) to give $\rho (\vmeanb \cdot \boldnabla ) \vmeanb %
= \tensor{\bm{M}} \cdot \partial \rho/\partial\xvec$. Hence, the leading order
contribution
to the total derivative does not arise from the explicit time derivative.

For completeness we also present the approach to steady-state values of the
particle-velocity covariances
\begin{subequations}
\begin{align}
\lim_{t \rightarrow \infty} \, \overline{v_x^{\prime} v_x^{\prime}} & =
\frac{k_B T}{m} \left [ 1 + \Stk^2 \left ( \frac{1}{2} -
\frac{9}{2 \beta t} \right ) \right ]
\Comma \\
\lim_{t \rightarrow \infty} \, \overline{v_x^{\prime} v_y^{\prime}} & =
- \frac{k_B T}{2 m} \, \Stk \left ( 1 - \frac{6}{\beta t} \right )
\Comma \\
\lim_{t \rightarrow \infty} \, \overline{v_y^{\prime} v_y^{\prime}} & =
\frac{k_B T}{m} \left ( 1 - \frac{2}{\beta t} \right )
\Teleia
\end{align}
\end{subequations}

The evaluation of the streamwise diffusion coefficient
$D_{xx}$ via the Green-Kubo relations Eq.~(\ref{eq:Dinfty})
requires the nonstationary correlation function
\beq
\langle v_x(t -\tau) y(t) \rangle = 2 \frac{k_B T}{m} \, \Stk
\left ( t - \tau \frac{2}{\beta} - \frac{1}{4 \beta} e^{-\beta \tau}
\right
) \quad \textrm{for} \quad \tau \geq 0 \quad .
\label{eq:DxxGK}
\eeq


\begin{thebibliography}{99}

\bibitem{ronis84}         D.\ Ronis,
\newblock                 Phys.\ Rev.\ A \textbf{29}, 1453 (1984).

\bibitem{hess84}          S.\ Hess and J.\ C.\ Rainwater,
\newblock                 J.\ Chem.\ Phys.\ \textbf{80}, 1295 (1984).

\bibitem{ramshaw79}       J.\ D.\ Ramshaw,
\newblock                 Phys.\ Fluids \textbf{22}, 1595 (1979).

\bibitem{delaMora82}      J.\ Fern\'andez de la Mora and D.\ E.\ Rosner,
\newblock                 J.\ Fluid Mech.\ \textbf{125}, 379 (1982).

\bibitem{brady}           G.\ Subramanian  and J.\ F.\ Brady,
\newblock                 Physica A \textbf{334}, 343 (2004).

\bibitem{rubi01}          I.\ Santamar\'{i}a-Holek, D.\ Reguera, and J.\ M.\ Rub\'{i},
\newblock                 Phys.\ Rev.\ E \textbf{63}, 051106 (2001).

\bibitem{delaMorapra82b}  J.\ Fern\'andez de la Mora and J.\ M.\ Mercer,
\newblock                 Phys.\ Rev.\ A \textbf{26}, 2178 (1982).

\bibitem{dufty83}         R.\ F.\ Rodr\'{i}guez, E.\ Salinas-Rodr\'{i}guez, and
                          J.\ W.\ Dufty,
\newblock                 J.\ Stat.\ Phys.\ \textbf{32}, 279 (1983).

\bibitem{sanmiguel}       M.\ San Miguel and J.\ M.\ Sancho,
\newblock                 Physica \textbf{99 A}, 357 (1979).

\bibitem{mike92}          M.\ W.\ Reeks,
\newblock                 Phys.\ Fluids A \textbf{4}, 1290 (1992).

\bibitem{pope}            D.\ C.\ Haworth and S.\ B.\ Pope,
\newblock                 Phys.\ Fluids \textbf{29}, 387 (1986).

\bibitem{eckhardt03}      B.\ Eckhardt and R.\ Pandit,
\newblock                 nlin.CD/0306016 (2003).

\bibitem{maxey83}         M.\ R.\ Maxey and J.\ J.\ Riley,
\newblock                 Phys.\ Fluids \textbf{26}, 883 (1983).

\bibitem{mike88}          M.\ W.\ Reeks,
\newblock                 Phys.\ Fluids \textbf{31}, 1314 (1988).

\bibitem{vandeven_jfm80}  R.\ T.\ Foister and T.\ G.\ M.\ van de Ven,
\newblock                 J.\ Fluid Mech.\ \textbf{96}, 105 (1980).

\bibitem{colloids}        J.\ K.\ G.\ Dhont,
\newblock                 \textit{An introduction to dynamics of colloids}
                          (Elsevier, Amsterdam, 1996).

\bibitem{mike_jfm04}      M.\ W.\ Reeks,
\newblock                 J.\ Fluid Mech.\ \textbf{522}, 263 (2005).

\bibitem{keith_thesis}    K.\ E.\ Hyland,
\newblock                 Ph.D. thesis, University of Strathclyde (1994).

\bibitem{keith99}         K.\ E.\ Hyland, S.\ McKee, and M.\ W.\ Reeks,
\newblock                 J.\ Phys.\ A: Math.\ Gen.\textbf{32}, 6169 (1999).

\bibitem{swailes97}       D.\ C.\ Swailes and K.\ F.\ F.\ Darbyshire,
\newblock                 Physica A \textbf{242}, 38 (1997).

\bibitem{resibois}        P.\ R\'{e}sibois and M.\ DeLeener,
\newblock                 \textit{Classical kinetic theory of fluids}
                          (John Wiley \& Sons, New York, 1977).

\bibitem{delaMorapra82a}  J.\ Fern\'andez de la Mora,
\newblock                 Phys.\ Rev.\ A \textbf{25}, 1108 (1982).

\bibitem{konstandopoulos90} A.\ G.\ Konstandopoulos,
\newblock                   J.\ Aerosol Sci.\ \textbf{21}, 983 (1990).

\bibitem{hess83}           S.\ Hess,
\newblock                  Physica \textbf{118 A}, 79 (1983).

\bibitem{ramshaw81}        J.\ D.\ Ramshaw,
\newblock                  Phys.\ Fluids \textbf{24}, 1210 (1981).

\bibitem{mathematica}      S.\ Wolfram,
\newblock                  \textit{The {M}athematica {B}ook}
                           (Wolfram Media/Cambridge University Press,
                           Cambridge (UK), 1999).

\bibitem{reguera03}        D.\ Reguera and J.\ M.\ Rub\'{i},
\newblock                  J.\ Chem.\ Phys.\ \textbf{119}, 9888 (2003).

\end{thebibliography}

\end{document}